\documentclass[doublecol]{epl2} 
\usepackage{mathrsfs}

\title{Signatures of two-step impurity mediated vortex lattice melting in Bose-Einstein Condensates}
\shorttitle{Signatures of two-step impurity mediated vortex lattice melting in Bose-Einstein Condensates} 

\author{Thudiyangal Mithun\inst{1,2} \and Somesh Chandra Ganguli\inst{2} \and Pratap Raychaudhuri\inst{2} \and Bishwajyoti Dey\inst{1}}
\shortauthor{T. Mithun \etal}

\institute{                    
  \inst{1} Department of Physics, SP Pune University, Pune 411007, India\\
  \inst{2} Tata Institute of Fundamental Research, Mumbai 400005, India.
}

\abstract{
We simulate a rotating 2D BEC to study the melting of a vortex lattice in presence of random impurities. Impurities are introduced either through a 
protocol in which vortex lattice is produced in an impurity potential or first creating the vortex lattice in the absence of random pinning and 
then cranking up the (co-rotating) impurity potential. We find that for a fixed strength, pinning of vortices at randomly distributed impurities 
leads to the new states of vortex lattice. It is unearthed that the vortex lattice follow a two-step melting via loss of positional and orientational
order. Also, the comparisons between the states obtained in two protocols show that the vortex lattice states are metastable states when impurities are introduced after the formation of an ordered vortex lattice. We also show the existence of metastable states which depend on the history of how the vortex lattice is created.}

\begin{document}
\maketitle
\section{INTRODUCTION}
In the past few years, study of vortex and its lattice formation in the Bose-Einstein Condensates (BECs) has grown beyond mere interest to understand its superfluidic nature. The main reason is that the analogous studies with respect to the various fields can be conducted. A few examples of such studies are, super-lattice structure formation has been studied in BECs in order to imitate a solid-state system \cite{Riordan:2016}, vortex lattice melting in presence of random impurities is studied for mimicking a type-II superconductor \cite{Mithun:2016}, quantum turbulence resulting from the vortex line motion has been compared with that of classical fluids \cite{Neely:2013,Tsatsos:2016}. The flexibility of BECs to be an ideal model in these cases has been attributed to its purity and the accuracy in the measurements. Recently methods for the analysis and observation of disorder in BECs have been demonstrated experimentally \cite{Rakonjac:2016}.
       \begin{figure}[!phb] 
       \vspace{-0.6cm}
   \onefigure[width=7.5cm,height=4.0cm]{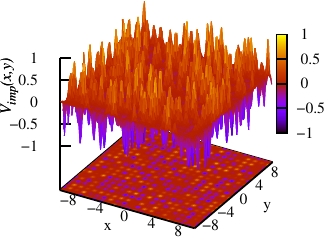}\hspace{-0.8em}
    \vspace{-0.9cm}
 \caption{\label{Fig. potential}{\footnotesize (Color online) (a) The random potential used for the problem. $V_{imp}(x,y)$ is created with a square optical lattice, where width of each Gaussian peak is 0.5 and distance between each peaks is 1, in units of $a_0$. Height of each peak has been changed with the help of random numbers which are uniformly distributed over $[-1,1]$ \cite{Mithun:2014}.}}
  \vspace{-0.4em}
  \end{figure}
        \begin{figure}[!htbp] 
      \vspace{-0.3cm}
     \onefigure[width=8.8cm,keepaspectratio]{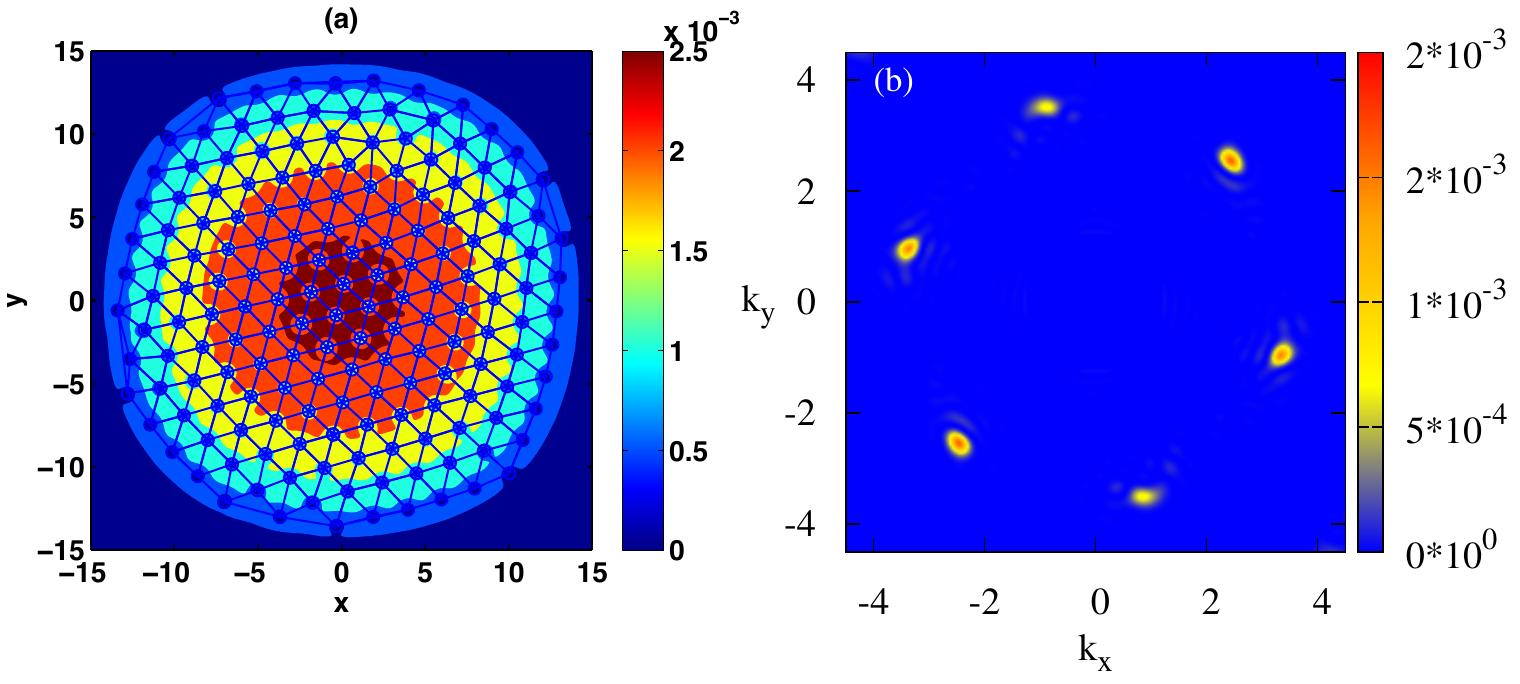}
    \vspace{-0.6cm}
 \caption{\label{Fig. delaunay}{\footnotesize (Color online) (a) Delaunay triangulated vortex lattice shows a perfectly ordered lattice with six-fold symmetry, and (b) shows the SF profile for the corresponding vortex lattice. Here $\Omega=0.9$.}}
    \vspace{-0.6cm}
  \end{figure}

 Even though the vortex lattice physics have been studied for last many years, the effect of impurity pinning potential on the vortex dynamics was not explored until very recently \cite{Mithun:2016}.  One can introduce impurity in BEC in a controlled way by laser speckle method to study the effect of random impurities in various observed physical phenomena. A few examples are, Anderson localization \cite{Billy:2008}, superfluid behavior \cite{Konenberg:2015} and superfluid-Mott-insulator transition \cite{Greiner:2012}. Rotating BECs with impurities provides a system where it is possible to display, in a controlled way, the interplay between interaction and disorder on the vortex dynamics. This competition is responsible for the impurity induced vortex lattice melting in BEC \cite{Mithun:2016}. The understanding of melting of the two-dimensional systems has an impact across several research fields \cite{Guillamon:2009} and vortex lattices in BECs are considered as ideal systems where the mechanism of such processes can be investigated. Also, the investigations related to the melting of vortex lattice in BECs are a subject of great interest because such problems are directly related to melting of Abrikosov lattice in Type-II superconductors, where melting process has been studied extensively to explore its role on the critical current \cite{Pastoriza:1994,Kierfeld:2000}. Even though the superfluids (BEC and liquid helium) and superconductors are very different systems,  the vortex dynamics in these systems show similar behaviour. For example, the equilibrium vortex lattice configuration in both the rotating BEC and type-II superconductors in applied magnetic field is same, the Abrikosov lattice \cite{Matveenko:2009}. Similarly, the dynamics of vortices in presence of an impurity (pinning of vortices) in both the systems also show similar behaviour. In presence of an impurity a vortex in type-II superconductor moves in a spiral path till it is pinned by the impurity \cite{Kato:1991}. We have earlier shown that a similar spiral path is also followed by a vortex in rotating BEC in presence of an impurity \cite{Mithun:2016}. It has been shown that the similarity in the vortex dynamics between the superfluid and the type-II superconductors is due to the fact that the transverse force or the Magnus force acting on a quantized vortex in a superfluid and type-II superconductor has the same form $\textbf{f} = \rho \textbf{K} \times \textbf{V}$ where $\rho$ is the mass density, $\textbf{K}$ is the quantized circulation vector and $\textbf{V}$ is the vortex velocity. Details are given in references \cite{Thouless:1996}. In addition to this, same kind of logarithmic interaction occurs between the vortices in both these systems \cite{Pitaevskii:2003}. The inhomogeneity introduced by the harmonic trap in BEC do not change the results. This is easily checked by studying the vortex dynamics in BEC for different trap size. Studying the problem of impurity induced vortex lattice melting in type-II superconductor is difficult because of the presence of natural impurities in the system. Therefore BEC provides a system where it is possible to simulate in a controlled way the interplay between interaction and disorder on the vortex dynamics. 

Our recent investigations on a two-Dimensional (2D) vortex lattice in BECs revealed that a vortex lattice can melt in presence of the random impurities \cite{Mithun:2016}. Such melting is fundamentally different from the more conventional thermal melting. It was originally proposed \cite{Vinokur:1998,Kierfeld:2000} in the context of vortex-matter in high-temperature superconductors. Here it is shown that the transition can be driven  by point disorder rather than temperature  even when thermal fluctuations alone is not sufficient to induce a transition to disordered vortex lattice. Recently, similar notions have been used by Ganguli et al \cite{Somesh:2016:a} and  Tsiok et al \cite{Tsiok:2017} among others. By disordered lattice we mean loss of long-range or translational periodicity  as well as orientational order. In absence of disorder the vortex lattice in rotating BEC is a regular triangular Abrikosov lattice and in presence of disorder the vortex lattice melts and the translational periodicity as well as orientational order of the regular vortex lattice is lost. In addition to the density profile ($|\psi ({\mathbf r})|^2$), the structure factor (SF) profile defined as $S({\mathbf k}) = \int d{\mathbf r} |\psi ({\mathbf r})|^2 e^{{\mathbf k.r}}$ also provide information about the translational and orientational order of the vortex lattice. In absence of impurities or disorder the SF profile shows the hexagonal lattice structure with sharp six periodic Bragg's peaks  \cite{Mithun:2016}. In presence of disorder the translational and orientational order of the vortex lattice is lost which is reflected in the corresponding SF profile where there are no periodic peaks and no regular hexagonal structure. It may be mentioned here that the true long-range translational order does not exist in a 2D materials. Further, in the context of vortex lattice in BEC, there will be quasi-long range order considering the finite size and non-uniform profile of the system due to the trapping potential. Further quantitative  information about the sequence of disordering of the vortex lattice is obtained from the calculation of orientational correlation function {\tt g}$_6(r)$  and the positional or spatial correlation function {\tt g}$_{\mathbf {K}}(r)$. Here {\tt g}$_6(r)$ is defined as {\tt g}$_6(r)=\left (\sum_{i,j}\Theta \left (\frac{\Delta r}{2} - \left |r - \left |{\mathbf r}_i - {\mathbf r}_j \right |\right |\right ) \cos 6 \left (\theta ({\mathbf r}_i) - \theta ({\mathbf r}_j)\right )\right )\times(1/n(r,\Delta r)$, where $\Theta (r)$ is the Heaviside step function, $\theta ({\mathbf r}_i) - \theta ({\mathbf r}_j)$ is the angle between the bonds located at ${\mathbf r}_i$ and the bond located at ${\mathbf r}_j$, $n(r,\Delta r) = \sum_{i,j}\Theta \left (\frac{\Delta r}{2} - \left |r -\left |{\mathbf r}_i - {\mathbf r}_j\right |\right |\right )$, $\Delta r$ defines a small window of the size of the pixel around $r$ and the sum is over all bonds. Similarly {\tt g}$_{\mathbf {K}}$ is defined as {\tt g}$_{\mathbf {K}}(r) = \left (\sum_{i,j}\Theta \left (\frac{\Delta r}{2} - \left |r - \left |{\mathbf R}_i - {\mathbf R}_j \right |\right |\right ) \cos {\mathbf K}.\left ({\mathbf R}_i - {\mathbf R}_j\right) \right )\times(1/N(r,\Delta r))$, where $\mathbf K$ is the reciprocal lattice vector obtained from the Fourier transform, ${\mathbf R}_i$ is the positon of the $i$-th vortex, $N(r,\Delta r) = \sum_{i,j}\Theta \left (\frac{\Delta r}{2} - \left |r -\left |{\mathbf R}_i - {\mathbf R}_j\right |\right |\right )$ and the sum is over all lattice points \cite{Somesh:2016:a}. The melting process is a gradual crossover from an ordered to a more disordered vortex arrangement with continuous increase of rotational frequency. If the disorder is removed, then the system goes back to the Abrikosov triangular lattice with long range order or translational periodicity. In Type-II superconductors such impurity mediated melting has been predicted due to the presence of crystalline imperfections in the crystalline lattice \cite{Vinokur:1998,Kierfeld:2000}, and recently, 
it has been found that these can lead to a two-step melting in a 3D system \cite{Somesh:2016:a}. In the first step, the number of dislocations increases with increase of rotational frequency. The positional order disappears but the orientational order is retained. While the second step involves creation of dislocations as well as disclinations with increasing rotational frequency and loss of both positional and orientational order. Similar two-step melting has also been observed in 2-D superconducting films \cite{Guillamon:2009}. While, in Ref. \cite{Guillamon:2009} the two-step melting was attributed to Berezinski-Kosterlitz-Thouless-Halperin-Nelson-Young (BKTHNY) transition\cite{Kosterlitz:1973,Halperin:1978,Nelson:1979,Pauchard:2006,Guillamon:2009}, the role of pinning was not explored in detail. Impurity mediated order-disorder transition has also been observed in 2-D colloidal crystals \cite{Pertsinidis:2008}. 
\begin{figure*}[!htbp] 
   \includegraphics[width=4.3cm,height=4.3cm,clip]{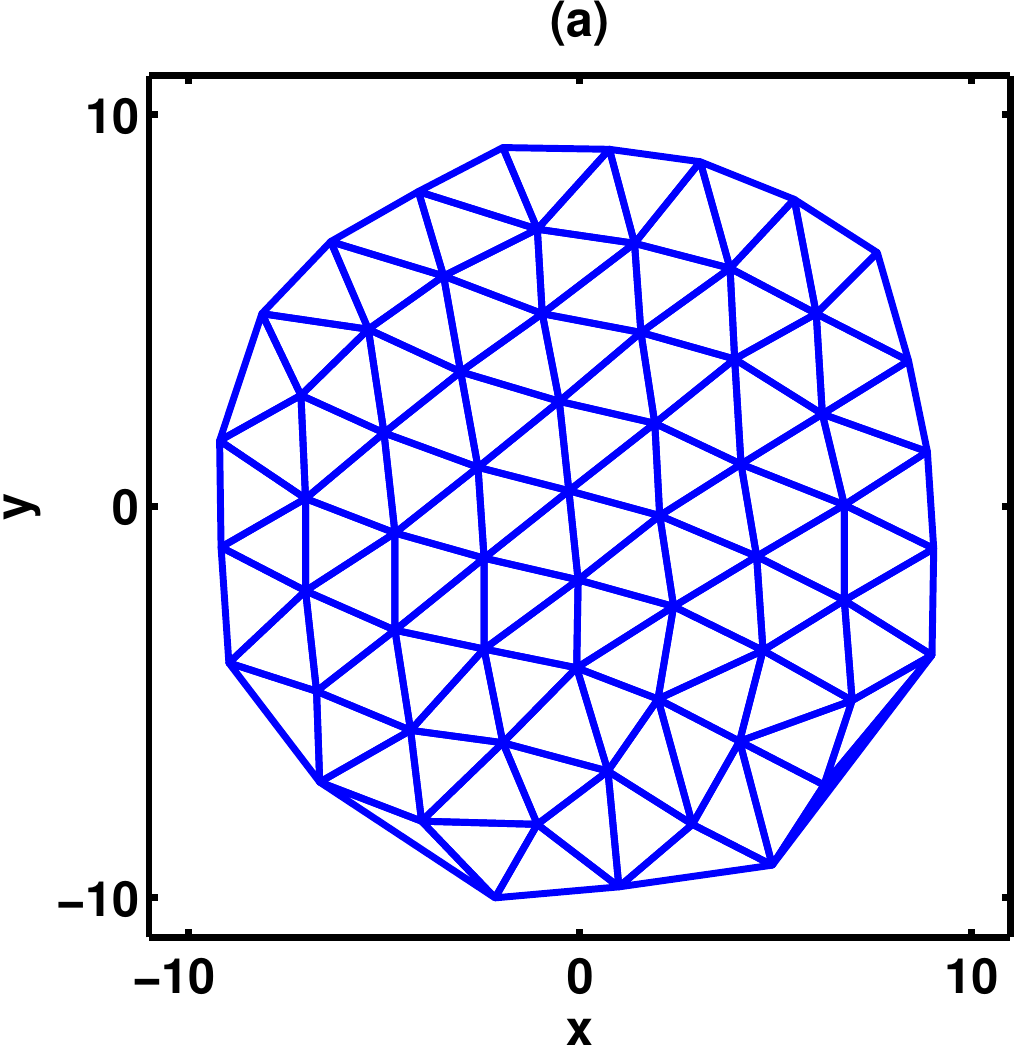}
   \includegraphics[width=4.3cm,height=4.3cm,clip]{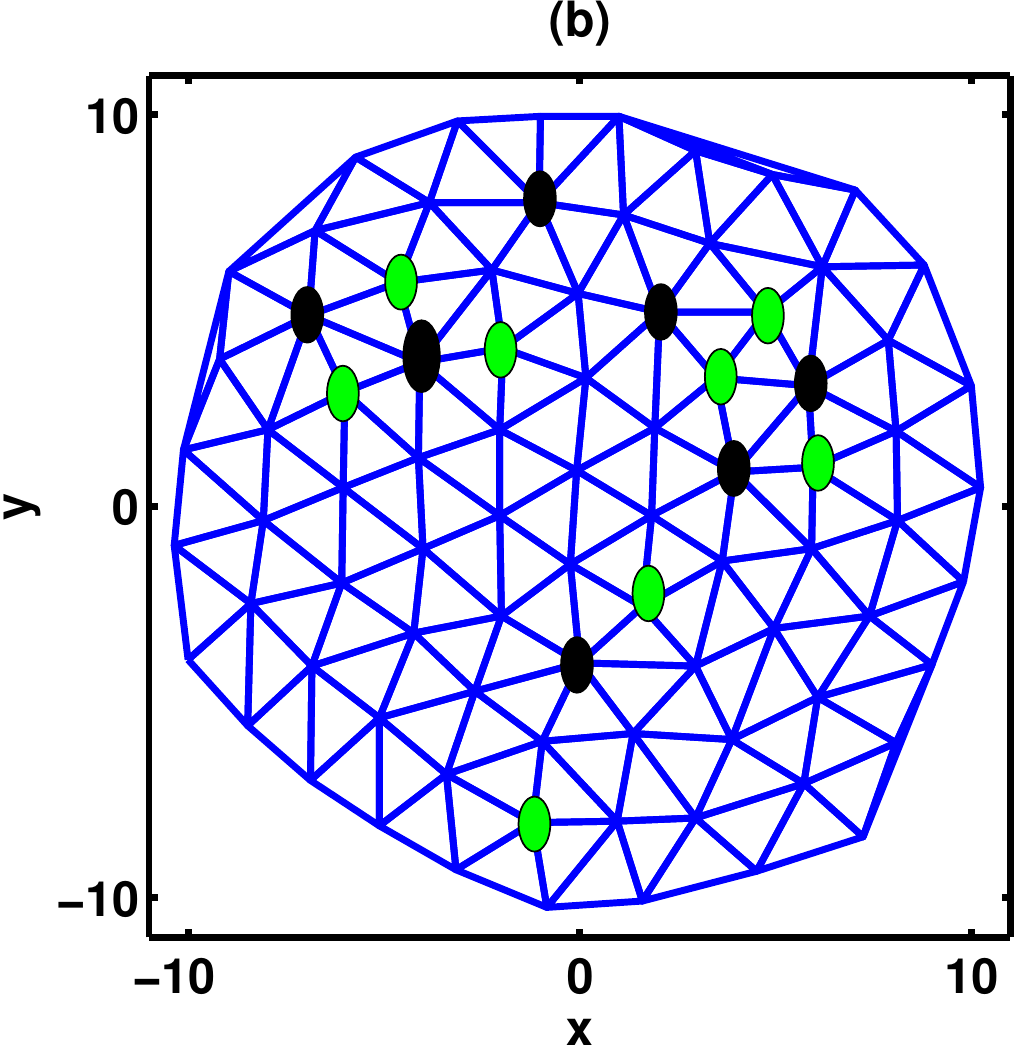}
   \includegraphics[width=4.3cm,height=4.3cm,clip]{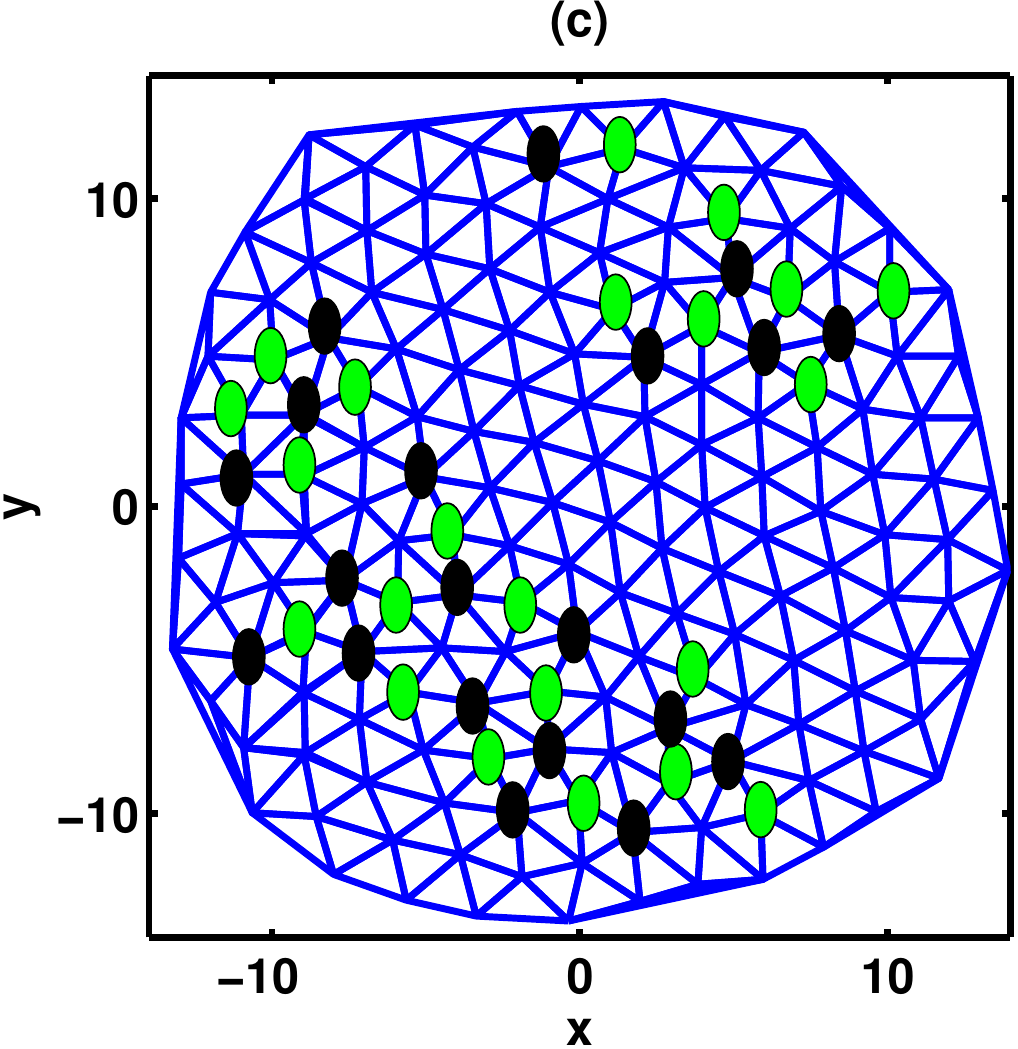}
   \includegraphics[width=4.4cm,height=4.4cm,clip]{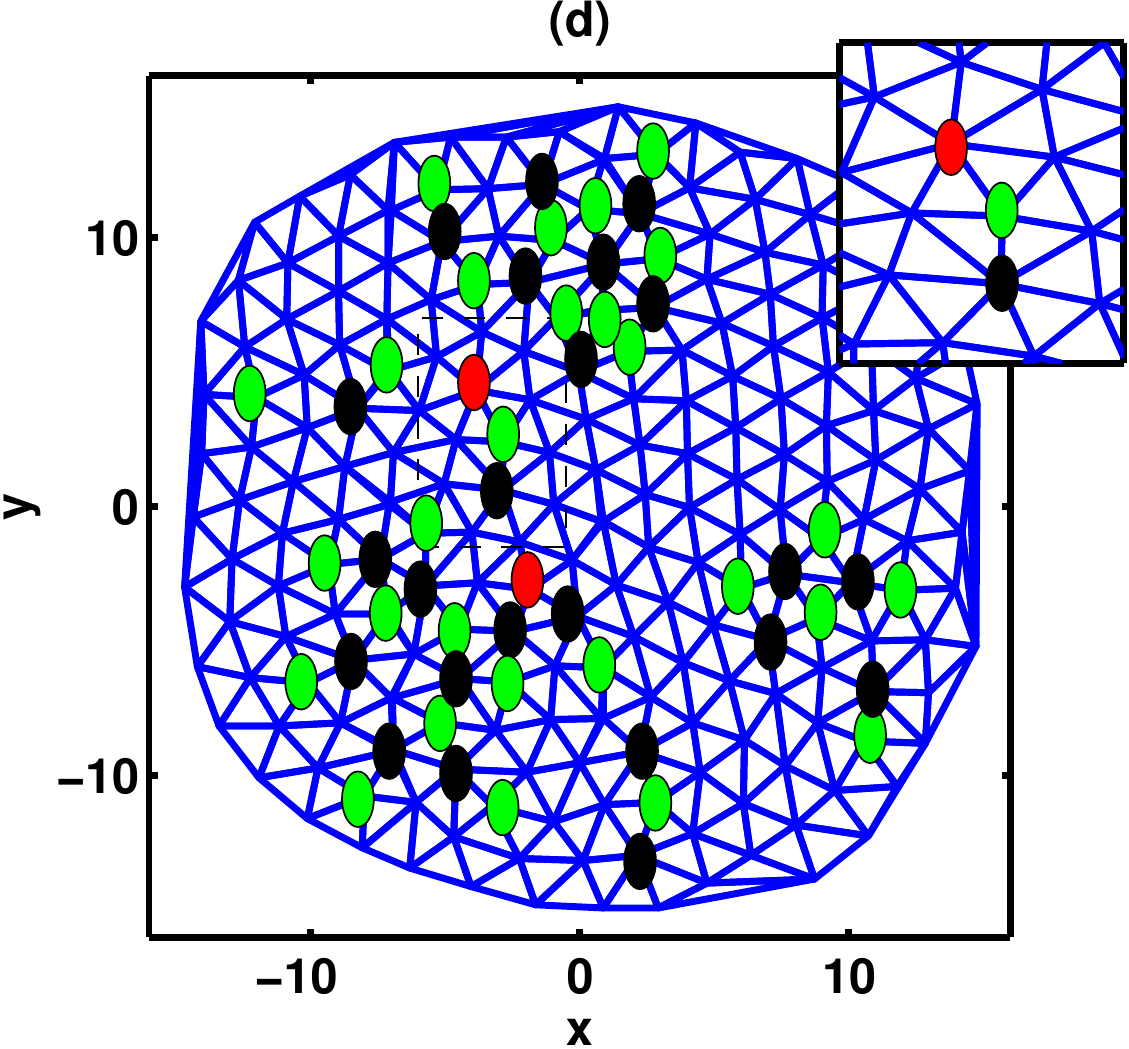}
    \vspace{-0.4cm}
 \caption{\label{Fig. zrs}{\footnotesize (Color online) Delaunay triangulated vortex profile showing the increase in the number of defects with rotation for 
ZRI at (a) $\Omega =0.6$, (b) $\Omega =0.7$, (c) $\Omega =0.90$ and (d) $\Omega =0.95$. Here, the five-fold and seven-fold dislocations are marked with green and black oval, respectively. Red ovals appearing in (d) shows the disclinations and the region is highlighted. Impurity strength is fixed at $V_0=2$.}}
     \vspace{-0.4cm}
 \end{figure*}
        \begin{figure}[!htbp] 
   \onefigure[width=8.5cm,keepaspectratio]{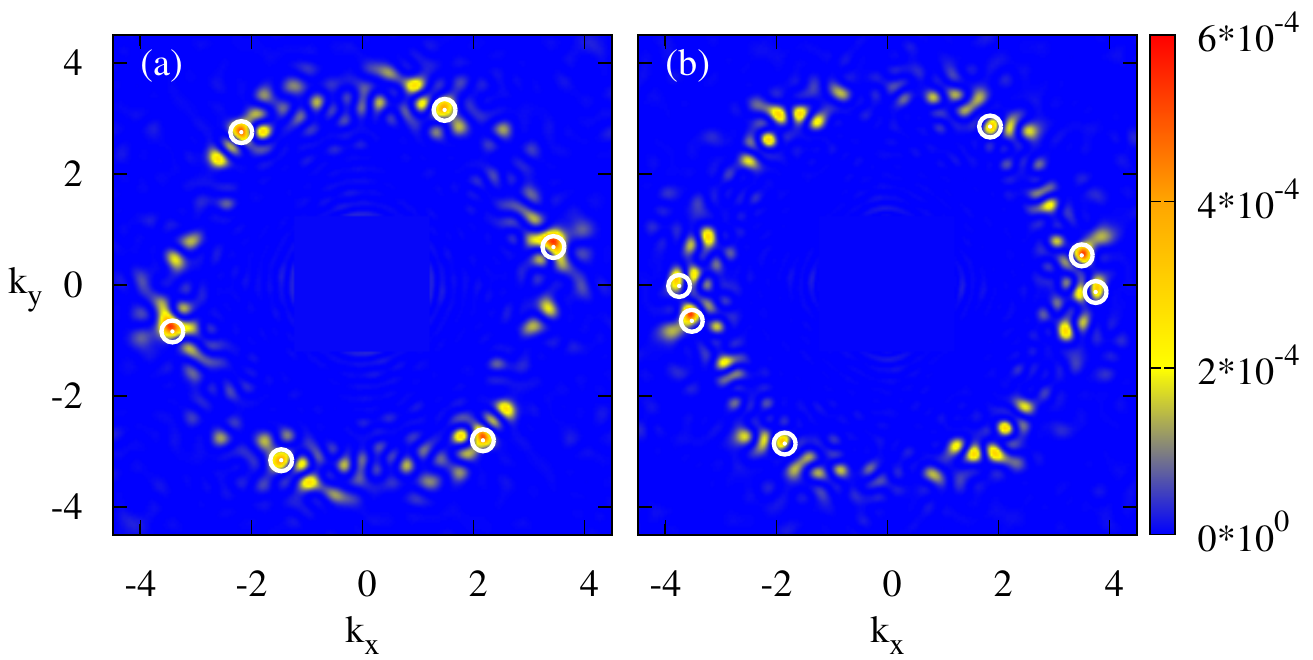}
    \vspace{-0.4cm}
 \caption{\label{Fig. zrs_sf}{\footnotesize (Color online) (a) and (b) are the SF profile corresponding to the Figs. \ref{Fig. zrs}(c) and (d), respectively. Here, (a) shows the orientational order by displaying six bright peaks, while (b) shows the loss of orientational order by the blurred structure. }}
    \vspace{-0.4cm}
  \end{figure}
  
In this paper we propose a BEC model, where one can study the vortex lattice melting analogous to those in weakly pinned Type-II superconductors.  We numerically investigate the melting of vortex lattice in a 2D BEC in presence of random impurities. Our results show that, in presence of a weak random pinning potential ($V_0 < \mu$), a 2D vortex lattice in BECs also follows a melting process, reminiscent of the two-step melting seen in \cite{Somesh:2016:a, Guillamon:2009}. It is shown that loss of long-range order occurs through two-steps with continuous increase of rotational frequency. At first the dislocations start to appear above a certain frequency and upon increase in frequency they gradually proliferate the lattice resulting in the loss of positional order, but the orientational order is retained. On further increase of the frequency the disclinations start to appear and consequently the orientational order of the vortex lattice is also lost. This effect is however absent in the absence of the random pinning landscape showing that it is fundamentally associated with the presence of random pinning.   We also show the existence of metastable states which depend on the history of how the vortex lattice is created. It is worth mentioning here that defects on the vortex lattice depends on the commensurate or incommensurate relation between the vortex lattice and the impurity potential \cite{Mithun:2016}. The spacing parameters of the impurity potential controls the  commensurate or incommensurate relation between the vortex lattice and the impurity potential.
  \section{THEORETICAL MODEL}
  It is well known that the 2D Gross-Pitaevskii equation (GPE)  can be obtained from the 3D GPE for a particular shape of the external harmonic trap potential. For the harmonic trap potential, $V_{trap} ({\mathbf r}) = \frac{1}{2}m[\omega_{r}(x^2 + y^2) +\omega_z z^2]$ the reduction happens when the aspect ratio of the external trap potential $\lambda_z = \frac{\omega_z}{\omega_{r}}> 1$ (pancake-shaped or oblate) or $\lambda_z < 1$ (cigar-shaped or prolate) \cite{Kasamatsu:2003}. We have considered $\lambda_z = 4$ as such potential causes axial alignment of the vortex. Also, it suppresses bending (Kelvin) modes caused by thermal fluctuations giving almost 2D vortex dynamics. Kelvin modes can be suppressed by tightening the confinement along the direction of the vortex line. Moreover, experiments have shown that the time scale of the vortex bending (longer than 1 sec) is much longer than the time scale of the dynamics of vortex lattice formation ($\sim$ 100msec) \cite{Rosenbusch:2002}. We, therefore consider our 2D analysis effective for the problem considered. Similar model of the vortex dynamics governed by the 2D GPE equation with aspect ratio $\lambda_z = 2.63$ has been used to explain more recent experimental observation of vortex lattice dynamics of BEC on square optical lattice \cite{Kato:2011, Williams:2010}. 
          \begin{figure*}[!htbp] 
     \includegraphics[width=5.5cm,keepaspectratio]{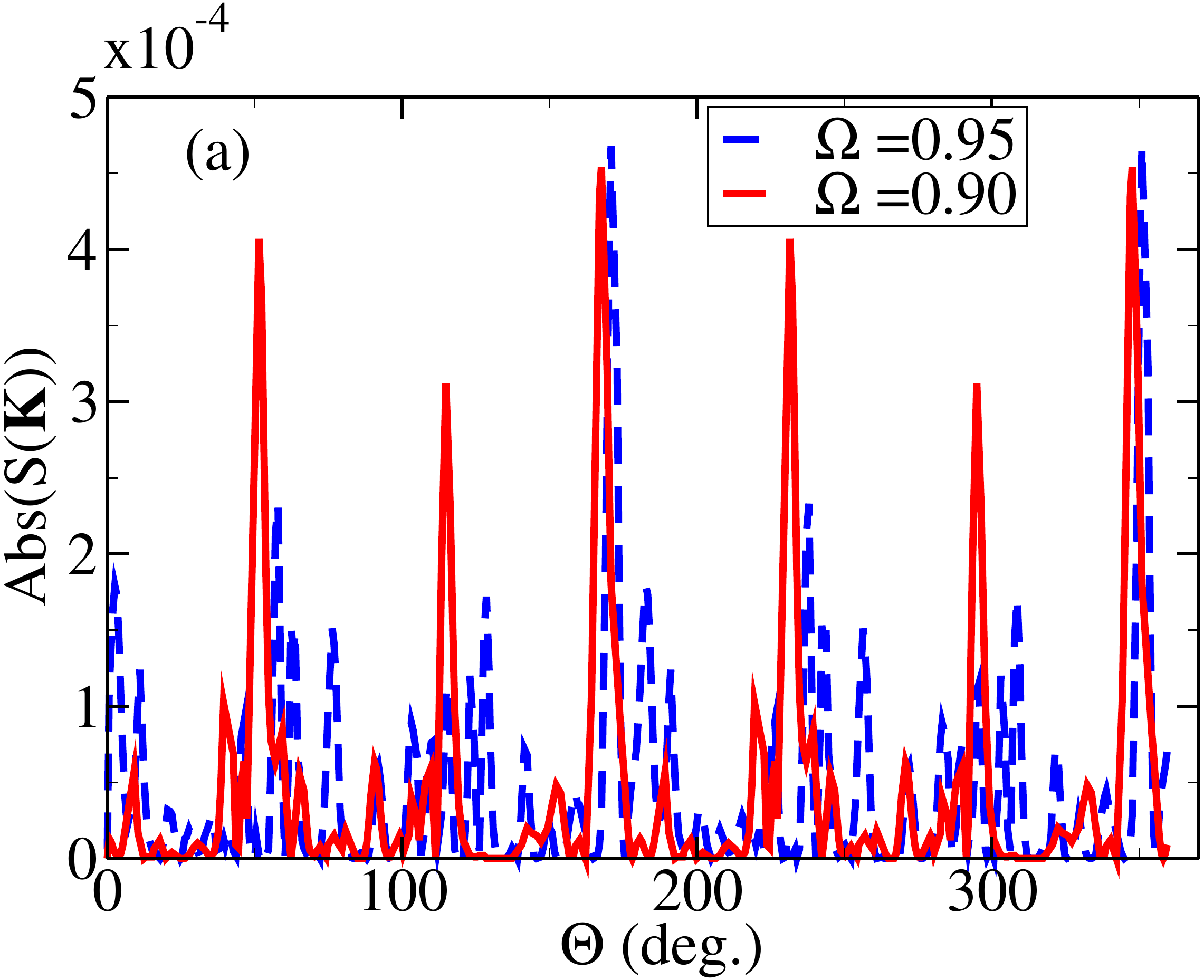}
      \includegraphics[width=6.0cm,height=4.0cm]{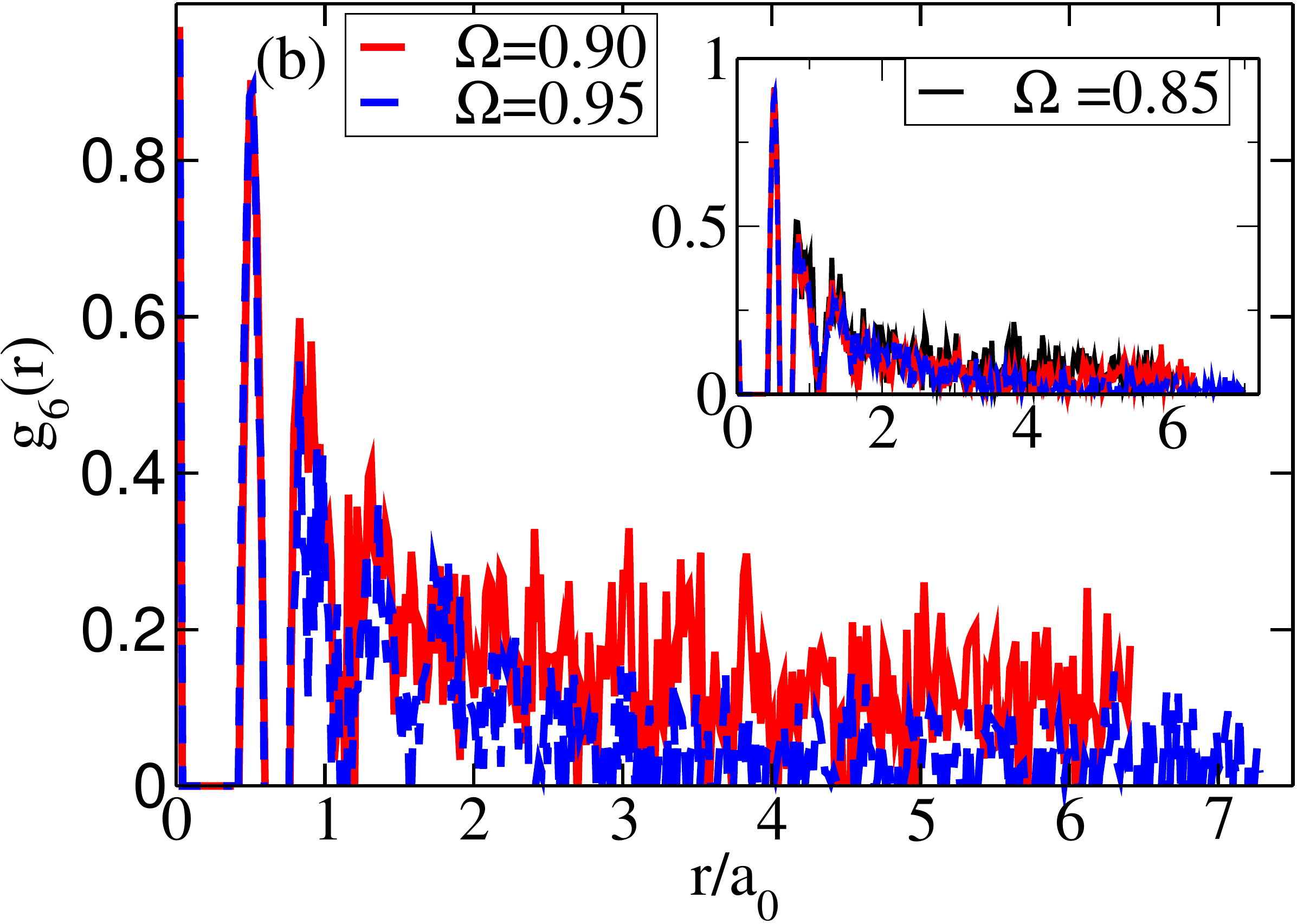}
   \includegraphics[width=5.8cm,height=4.0cm]{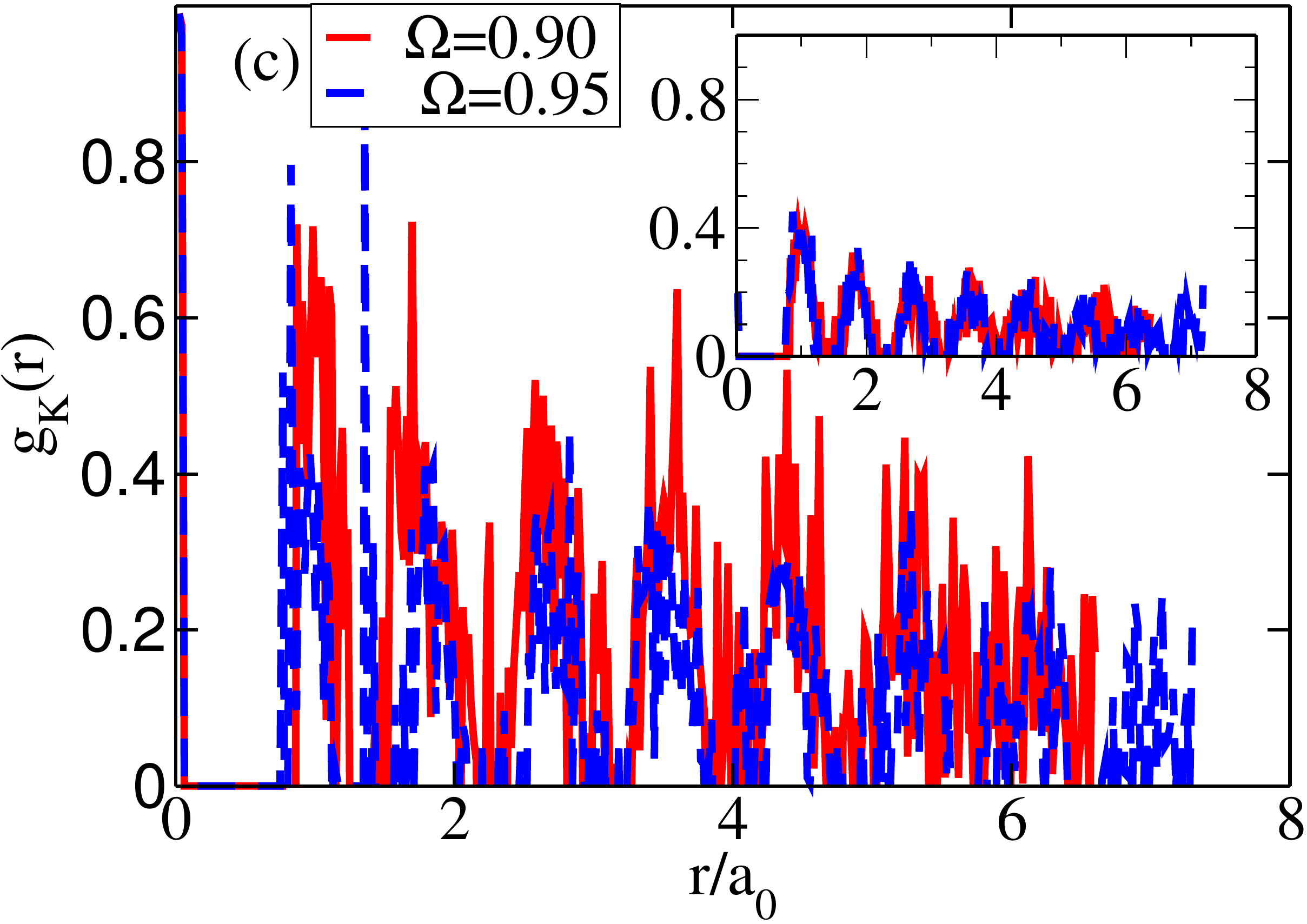}
    \vspace{-0.4cm}
 \caption{\label{Fig. Intensity}{\footnotesize (Color online) (a) Shows the intensity (Abs(S(\textbf{K})) as a function of the azimuthal angle for 
 $\Omega=0.95$ and $\Omega=0.90$ in the ZRI protocol. Here intensity data is calculated at the radial maximum. (b) Orientational correlation function, {\tt g}$_6$ and (c) and positional correlation function, {\tt g}$_K$ (averaged over the principal symmetry directions) as a function of $r/a_0$, where $a_0$ is the average lattice constant. The inset shows the correlation functions averaged over few random realizations.}}
    \vspace{-0.2cm}
  \end{figure*}
             \begin{figure*}[!htbp] 
   \includegraphics[width=4.4cm,height=4.4cm,clip]{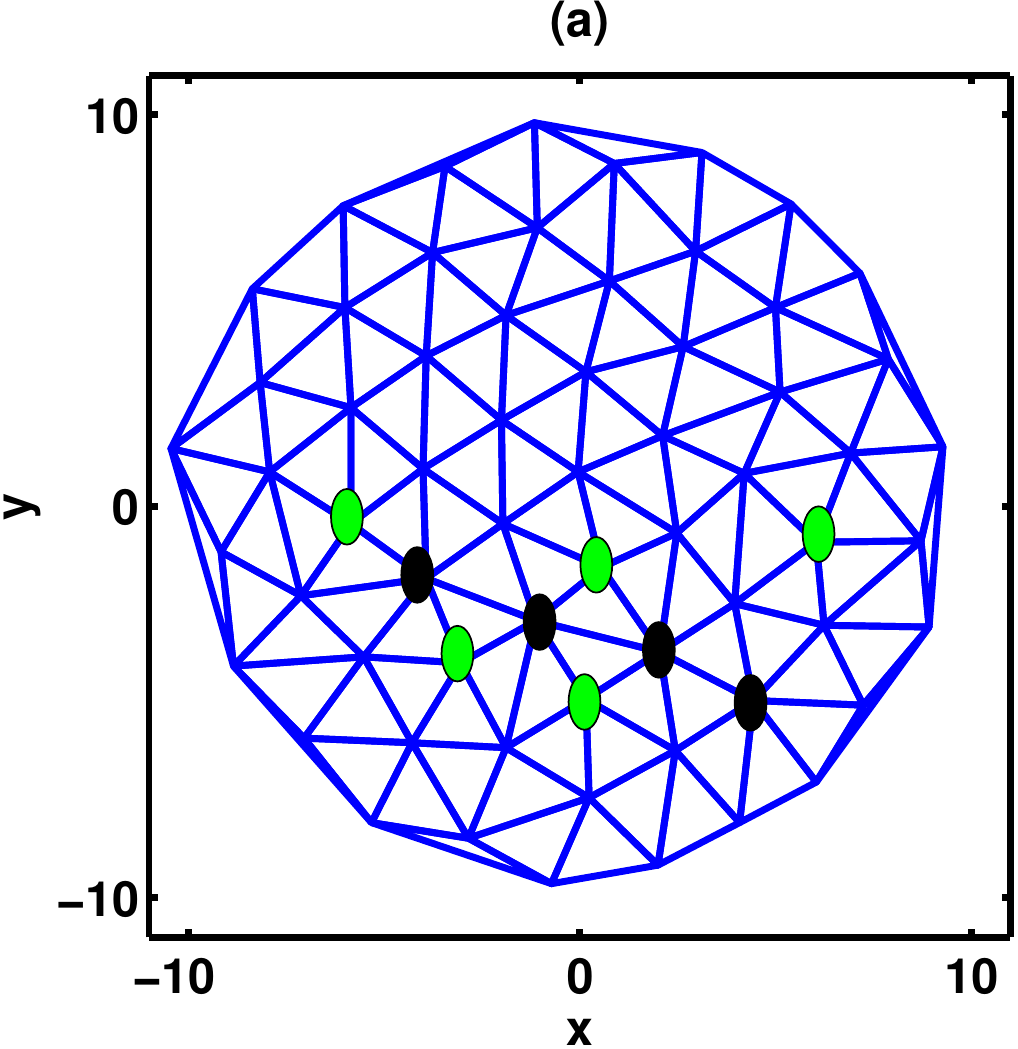}
   \includegraphics[width=4.4cm,height=4.4cm,clip]{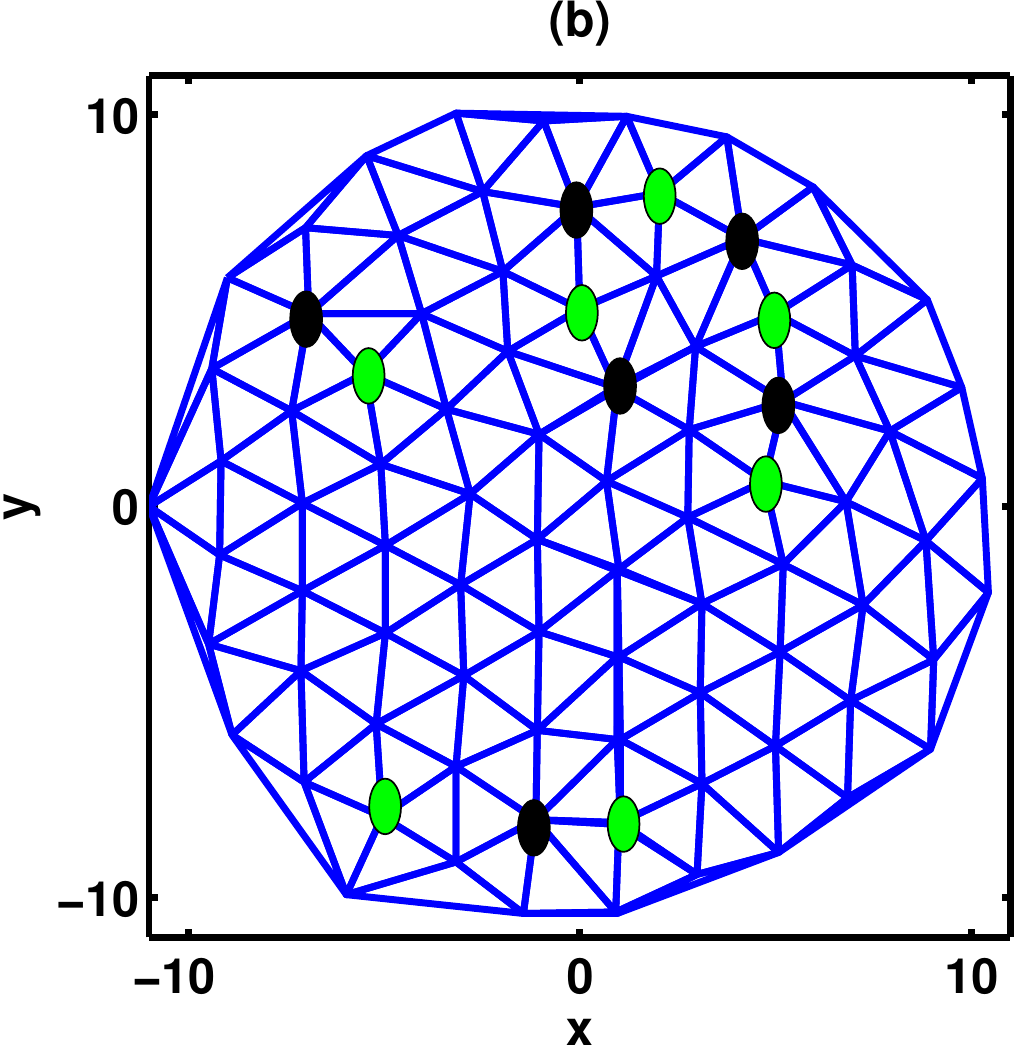}
   \includegraphics[width=4.4cm,height=4.4cm,clip]{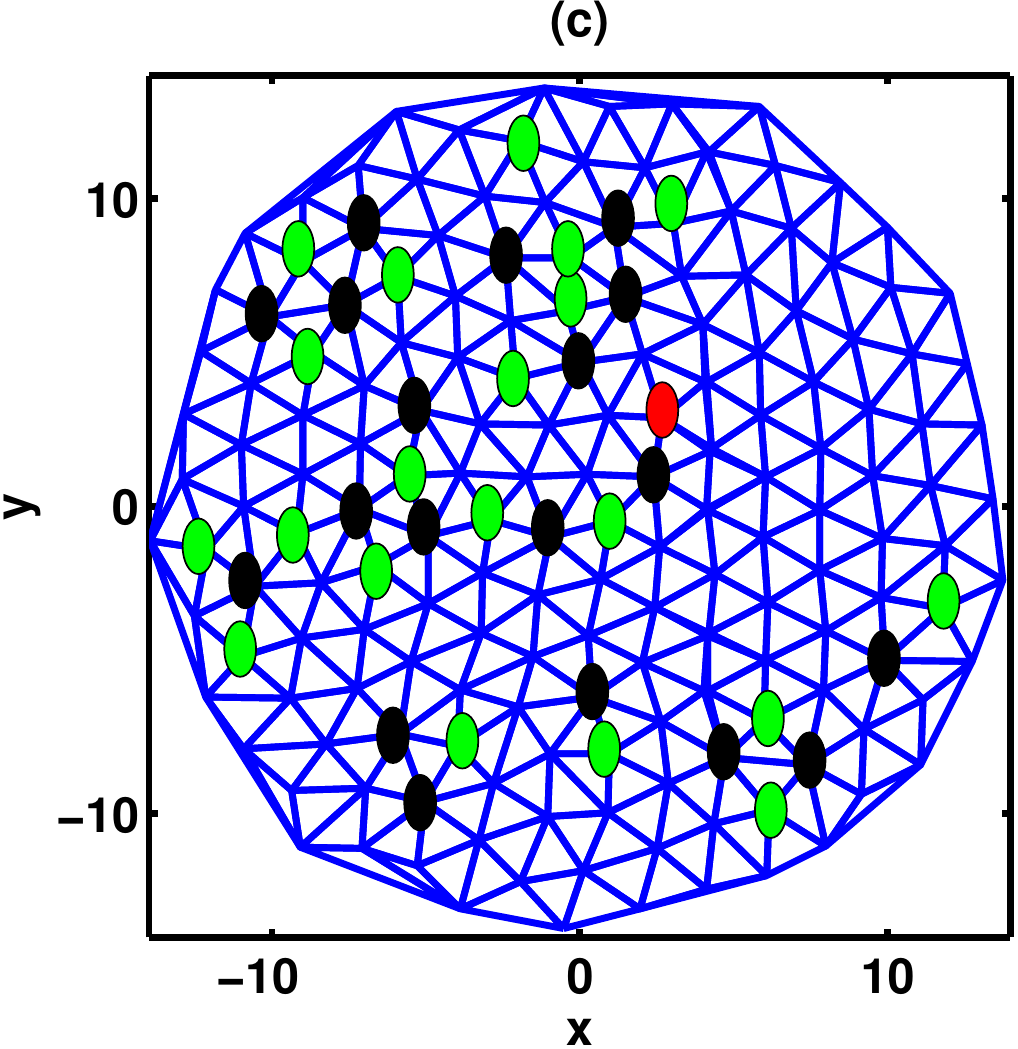}
   \includegraphics[width=4.4cm,height=4.4cm,clip]{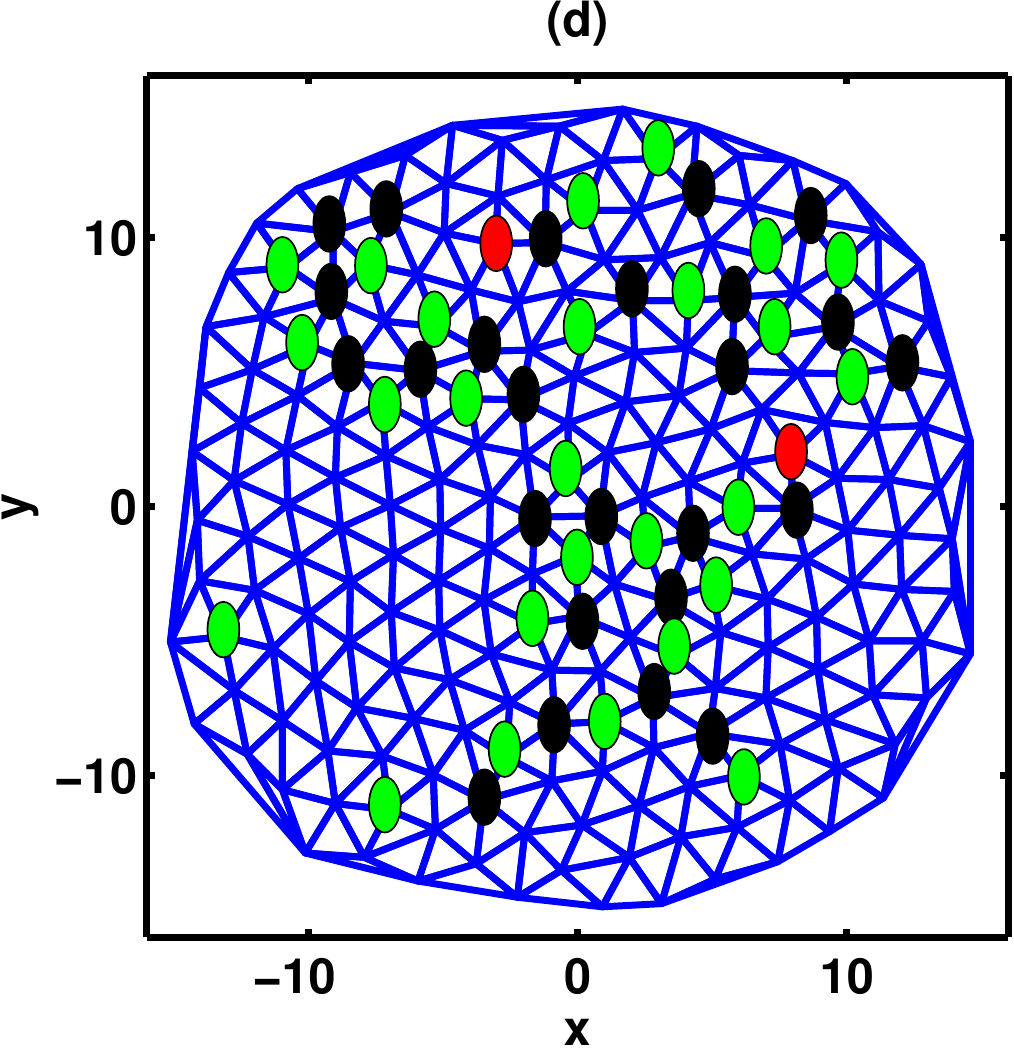}
       \vspace{-0.4cm}
 \caption{\label{Fig. rs}{\footnotesize (Color online) (a) Delaunay triangulated vortex profile showing the increase in the number of defects with rotation for RI at (a) $\Omega =0.6$, (b) $\Omega =0.70$, (c) $\Omega =0.90$, and (d) $\Omega =0.95$. Impurity strength is fixed at $V_0=2$.}}
      \vspace{-0.4cm}
  \end{figure*}
The two-dimensional (2D) dimensionless time-dependent GPE \cite{Kasamatsu:2003} is,
\begin{equation}
\label{eq.1}
\small  (i-\gamma)\psi_{t}=[-\frac{1}{2}(\partial_{x}^2+\partial_{y}^2 )+V(x,y)-\mu+p|\psi|^2-\Omega L_{z}]\psi \,
\end{equation}
where $V(x,y)=V_{trap}(x,y)+V_0 \times V_{imp}(x,y)$, $\Omega$ is the rotational frequency and $L_{z}$ is the angular momentum in z-direction. 
The spatial coordinates, time, rotational frequency, condensate wave function and energies are in units of $a_0$ , $\omega_{r}^{-1}$, $\omega_{r}$, 
$\sqrt{N}a_0^{-3/2}$ and $\hbar \omega_{r}$, respectively. Condensate is trapped in a harmonic trap potential, $V_{trap}=\frac{1}{2}(x^2+y^2)$. Here 
$p = \frac{4\pi a_s N}{a_0}\sqrt{\frac{\lambda_z}{2 \pi}}=11490$,  $N =1.5\ast10^6$, $a_s=4.76nm$, 
$a_0=\sqrt{\frac{\hbar}{m\omega_r}}=6.23\mu m$, 
$(\omega_r,\omega_z)=2\pi(3,12)$Hz. $\gamma$ is 
the dissipation factor  and is set to 0.03 in our simulation. As discussed in \cite{Mithun:2016},
the origin of the dissipation is due to the collision of the noncondensed atoms (presence of thermal component in the trap) with the condensate atoms.
The dissipation has been confirmed by the experiments  on trapped BEC \cite{Jin:1997, Mewes:1996:a}. In order to accommodate the dissipation Choi 
\textit{et al} \cite{Choi:1998}  modified the GPE by including the dissipative term to describe the dynamics of the condensate. 
They obtained the value, $\gamma = 0.03$ by fitting their theoretical results with the above mentioned experiments and thereafter the same value has 
been used for several studies  \cite{Mithun:2014, Tsubota:2002, Kasamatsu:2003, Kato:2011, Kasamatsu:2006}.  The dissipative term helps faster convergence to the stationary state. We have numerically checked that the results do not change even for $\gamma = 0$. It has been 
shown earlier that the phenomenological GPE (Eq. (1))  can also be derived from the generalized GPE at finite temperature \cite{ Kasamatsu:2003}.
Because of the dissipative term, the time development of the dynamical equation (Eq. (1)) neither conserve the norm of the wave function nor the energy. In order to conserve the norm of the wave function for nonzero value of $\gamma$, the chemical potential is treated as time dependent and is adjusted at each time step. In our simulation, we calculate the correction to chemical potential in every step as \cite{Mithun:2016, Kato:2011, Kasamatsu:2003, Tsubota:2002} $\Delta \mu = (\Delta t)^{-1} ln[\int d^2r |\psi(t)|^2/\int d^2r|\psi(t+\Delta t)|^2],$
 which conserves the norm and the total energy decreases monotonically eventually leading to the stationary states.   Here $V_{imp}(x,y)$ is the disorder potential (see Fig. \ref{Fig. potential}) and $V_0$ denotes its 
 strength. We have numerically checked that the results do not change even if we take repulsive impurity at all sites, i.e. choose the random numbers uniformly distributed over $[0,1]$. We have further checked that the results are valid for different realization of disorder potential and also for different values of the impurity strength $V_0$. We have shown that the correlation functions averaged over few random realization also shows the expected decay profile.
                   \begin{figure}[!htbp] 
       \vspace{0pt}
               \onefigure[width=8.7cm,keepaspectratio]{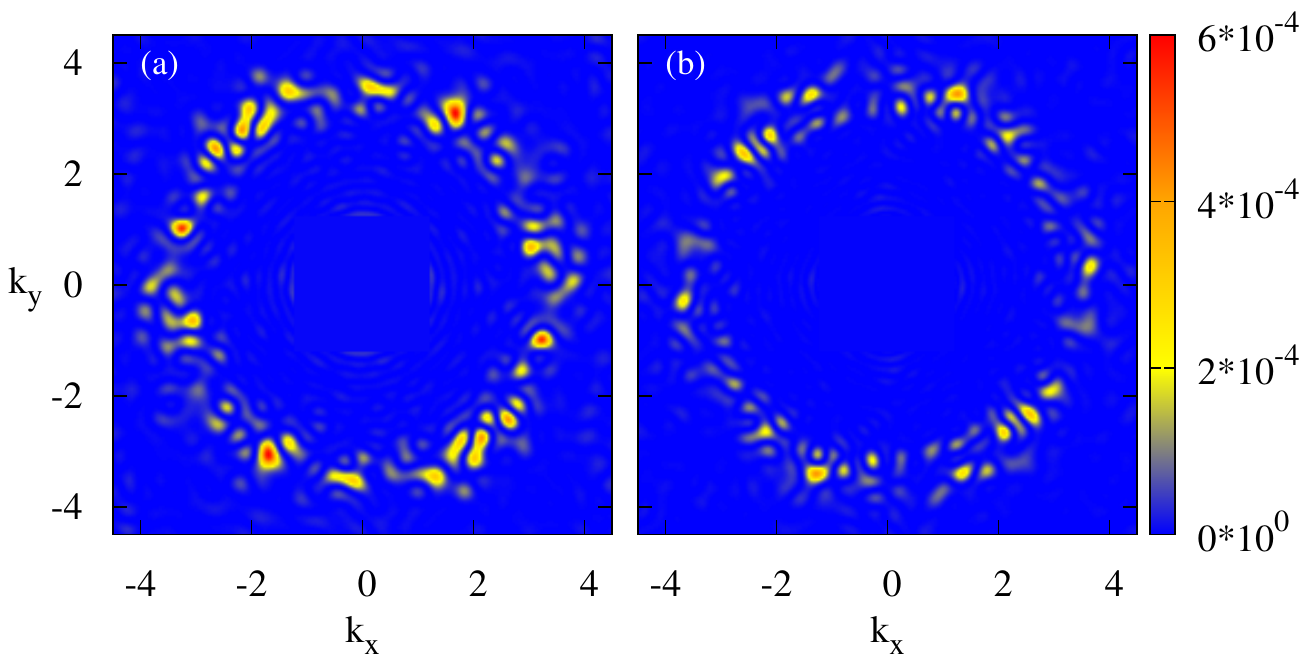}
    \vspace{-0.4cm}
 \caption{\label{Fig. rs_sf}{\footnotesize (Color online) SF profile showing the increase in the number of defects with rotation for RI at (a) $\Omega =0.9$ and (b) $\Omega =0.95$. Here, both (a) and (b) shows the loss of positional and orientational order.}}
    \vspace{-0.4cm}
  \end{figure} 

  In BECs the random impurity potential is created and controlled by laser speckle method \cite{Palencia:2010}. The vortex lattice in a BEC in presence of a random pinning potential can be created in two ways: i) either a vortex lattice can be created by rotating the BEC in a co-rotating impurity potential or by ii) first creating the vortex lattice in the absence of random pinning and then cranking up the (co-rotating) impurity potential. These two ways has obvious relation with the two commonly known protocols of creating a vortex lattice in a type-II superconductor: zero field cooling (ZFC) protocol and the field cooling (FC) protocol. The final effect of both these protocols is the difference in the way the vortices feels the effective impurity or pinning potential due to the presence of impurities in the superconductor. The relation can be understood as follows. The strength of the pinning or impurity potential in a superconductor is proportional to the condensation energy $(1/2)N(0)\Delta^2$, where $N(0)$ is the density of states at the Fermi level and $\Delta$ is the superconducting energy gap which increases monotonically from zero as the superconductor is cooled below $T_{\mathrm{c}}$. Therefore as the superconductor is cooled (in zero applied magnetic field) below $T_{\mathrm{c}}$  the pinning potential increases (as $\Delta$ increases) till it attains its maximum value (as $\Delta$ attains its maximum value). The magnetic field is then  applied to create vortex lattice after the pinning potential reaches its maximum value. Thus in the ZFC protocol the sample is first cooled below $T_c$ to create impurity or pinning potential and then the magnetic field is applied to create vortex lattice. The vortices feels the maximum value of the pinning or impurity potential.  On the other hand in the FC protocol the magnetic field is first applied to create vortex lattice  just below $T_c$ when the pinning potential is infinitely small, and then as the sample is cooled (in presence of the applied field) the vortex lattice adjusts to accommodate the increasing strength of random pinning potential. Thus in the FC protocol, first magnetic field is applied to create vortex lattice and then the sample is cooled in presence of the applied field.  In our BEC model the temperature is not a variable. As mentioned above, the effect of finite temperature is included phenomenologically through the damping parameter $\gamma$. The vortex lattice in BEC is created by rotating the condensate. So, rotation in BEC context is equivalent to application of magnetic field in type-II superconductor as both create vortex lattice  \cite{Matveenko:2009}. We can introduce impurity to the BEC either before the vortex lattice formation or after forming the vortex lattice. By following the same logic as superconductor, we first add the impurity potential and then rotate the condensate to generate the vortices in order to mimic a ZFC state. While, for mimicking FC state in BEC, we first rotate the condensate to create vortex lattice and then introduce the impurity or pinning potential.  Therefore, for differentiating these two ways in BEC context, we hereafter refer the protocols as (i) zero  rotation impurity (ZRI) and (ii)  rotation impurity (RI), respectively.       
         \begin{figure*}[!htbp] 
   \includegraphics[width=4.4cm,height=4.5cm,clip]{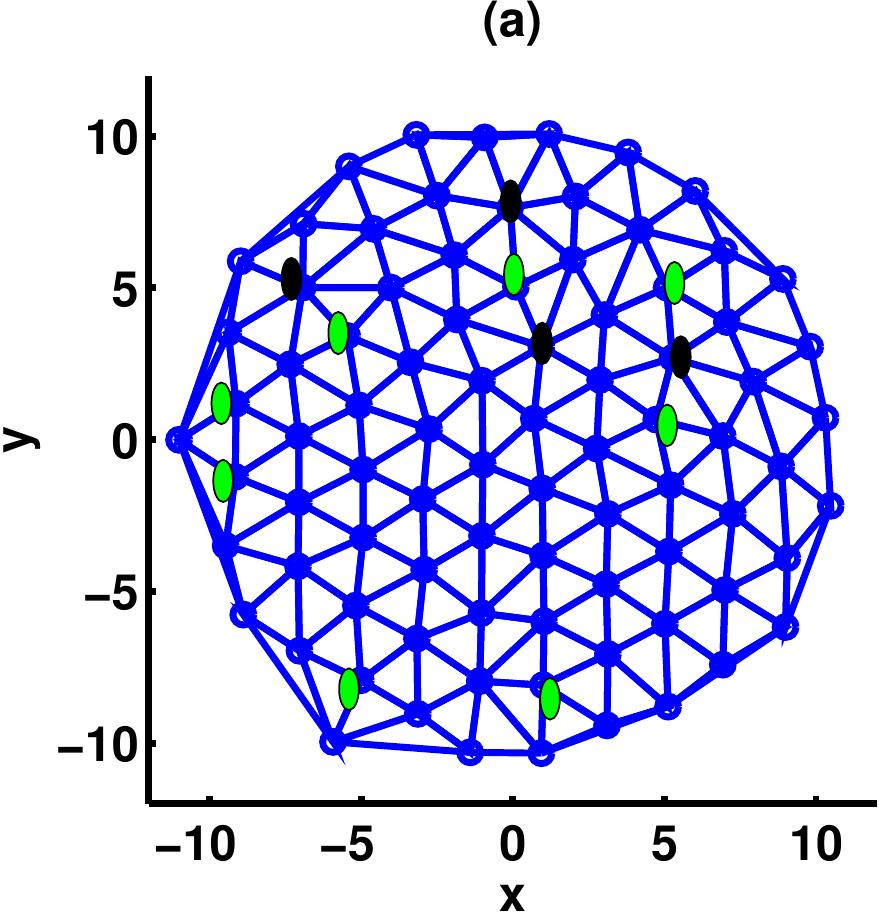}
   \includegraphics[width=4.4cm,height=4.5cm]{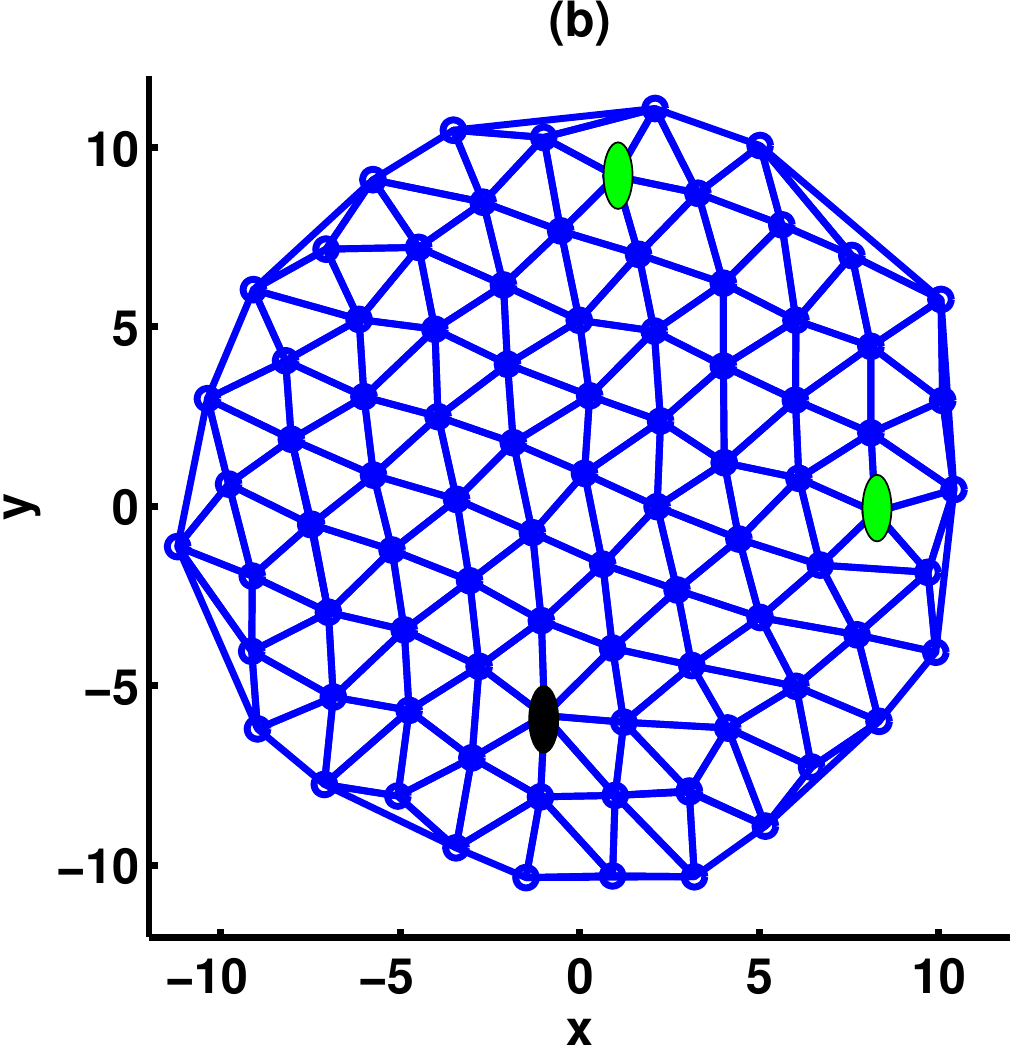}
   \includegraphics[width=4.4cm,height=4.5cm]{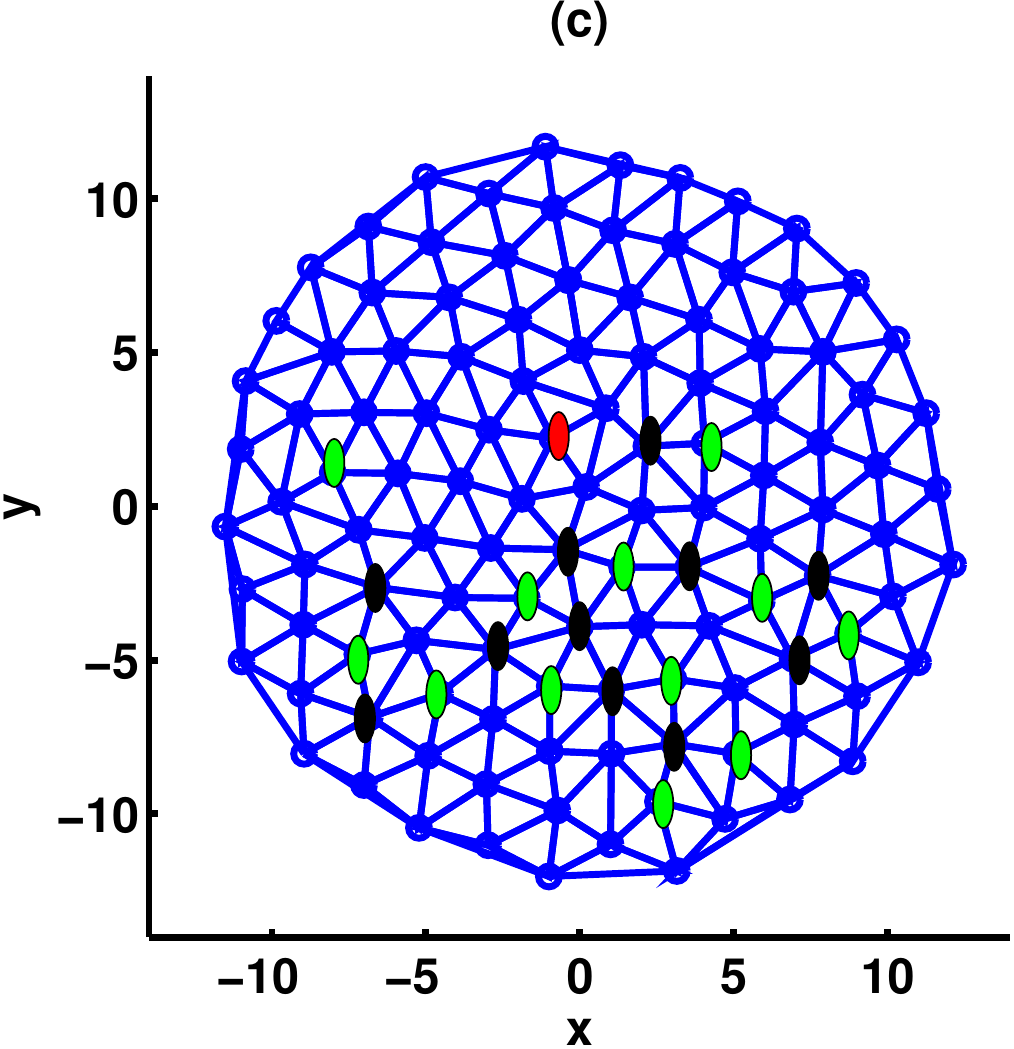}
    \includegraphics[width=4.4cm,height=4.5cm]{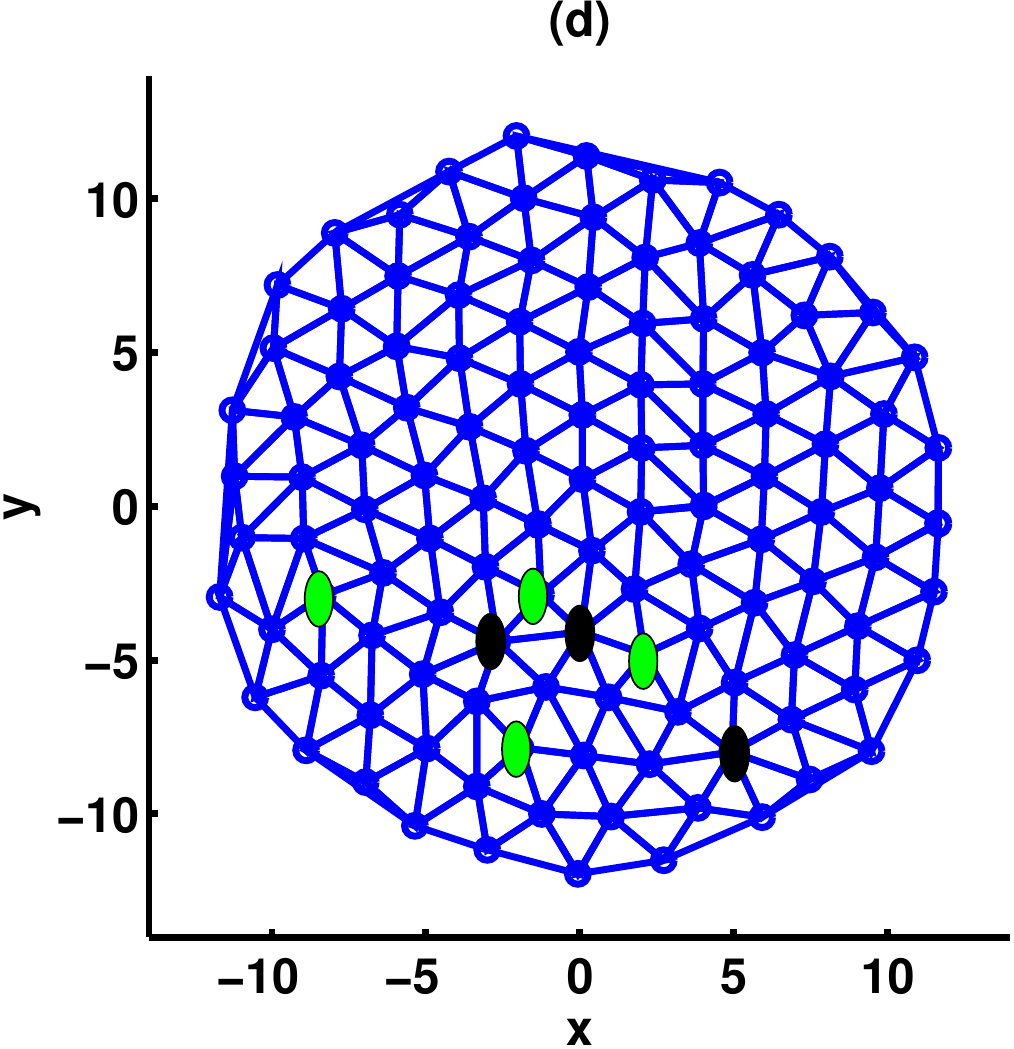} 
        \vspace{-0.8cm}
 \caption{\label{Fig. pulsing_ir}{\footnotesize (Color online) (a) and (c) show Delaunay triangulated disordered vortex lattice at $\Omega=0.7$ and $\Omega=0.8$, respectively  for RI case; while, (b) (for \ref{Fig. pulsing_ir}(a)) and (d) (for \ref{Fig. pulsing_ir}(c)) show much ordered lattice after perturbing it with higher rotational frequency. In both cases, initial perturbation involve increasing the rotation frequency by an additional amount 0.05 and then decreasing to their respective original frequencies. Later, the same process is repeated with an additional frequency 0.07.}}
    \vspace{-0.4cm}
  \end{figure*}
 \section{NUMERICAL ANALYSIS}
The time-splitting Crank-Nicolson method \cite{Muruganandam:2009, Young:2016} is used to solve the GPE numerically with the
spatial steps $\Delta x = \Delta y =0.04$. All simulations are done for 
 the region $[-24,24] \times [-24,24]$. In order to show that in the absence of an impurity potential we realize an ordered Abrikosov lattice up 
to high rotational frequencies, we plot $|\psi|^2$ in Fig. \ref{Fig. delaunay} for  $\Omega=0.9$. The center of vortices are detected by 
calculating the extrema of the superfluid vorticity $\omega=\nabla \times v_s$ \cite{Davis:2009}. After extracting
the location of vortices, the Delaunay triangulation is used to determine the number of the nearest neighbours. In this triangulation, first a 
triangle is considered by joining any three points in the lattice. These three points are nearest neighbours only if the circumcircle of this triangle does not contain any other lattice point.
The Delaunay triangulated vortex lattice shows six-fold nearest neighbour coordination for all vortices except for those at the boundary.  
The corresponding Fourier transform of the $|\psi|^2$ ($S(\textbf{K})=F\{|\psi|^2\}$) also shows the hexagonal structure with sharp six symmetric 
Bragg's peaks (Fig. \ref{Fig. delaunay}(b)). In the present work, the maximum number of vortices considered is around 230. Even for this number,
the higher value of filling factor (ratio of the number of particles to the number of vortices) greater than 5000 leaves the system in the mean-field 
quantum Hall regime. It is well above to the limit $\sim 10$ where vortex melting is expected due to the quantum fluctuations \cite{Ho:2001}. In this regime, GPE is able to describe the vortex lattice.  
\subsection{ZRI protocol}
  Now, we focus on the vortex lattice state in the presence of impurity. In the ZRI protocol, the competitive interactions arising from the
  vortex-vortex and vortex-impurity leads to vortex configurations as shown in Fig. \ref{Fig. zrs} for the impurity strength, $V_0 =2$. 
  Fig. \ref{Fig. zrs}(a) shows an ordered equilibrium state at low rotational speed, $\Omega = 0.6$. Increase in rotational speed to $\Omega =0.70$ 
  shows the appearance of dislocations as depicted in Fig. \ref{Fig. zrs}(b). Dislocation is a defect consisting of a pair of adjacent lattice points
  with five-fold and seven-fold coordination. On further increase of the rotational frequency to $\Omega =0.90$, Fig. \ref{Fig. zrs}(c) shows the 
  increase in number of dislocations. For $\Omega =0.95$, in addition to the dislocations, disclinations also appears in the vortex lattice as 
  in Fig. \ref{Fig. zrs}(d). Disclinations are isolated or unpaired five-fold or seven-fold coordinated lattice sites. The inset figure shows 
  that when a five-fold (green oval) coordination pairs with seven-fold (black oval), a seven-fold (red oval) is left without having an adjacent 
  five-fold to pair. For calculating the disclination, the unpaired coordination sites near to the boundary are not considered. Appearance of small
  peaks in the Fourier transform of the Fig. \ref{Fig. zrs}(c) (Fig. \ref{Fig. zrs_sf}(a)) reveals that the positional order started to decrease in 
  the system; while, the system still retains the orientational order by displaying six peaks. The six peaks are marked with white circles, where 
  first six peaks with maximum intensity are considered. The Fourier transform in Fig. \ref{Fig. zrs_sf}(b) corresponding to $\Omega =0.95$ case 
  shows that with the appearance of disclination, orientational order has also been destroyed by displaying the loss of six-fold symmetry. The loss 
  of positional and orientational orders are can also be verified by plotting intensity (Abs(S(\textbf{K})) as a function of the 
  azimuthal angle in Fig. \ref{Fig. Intensity}(a). For $\Omega =0.90$ case six clear peaks are visible: while, for $\Omega =0.95$ case these six 
  peaks have been destroyed.   It shows the loss of orientational order in $\Omega =0.95$ case. Similarly the decay of the correlation functions {\tt g}$_6(r)$ and {\tt g}$_K(r)$ (see \ref{Fig. Intensity}(b) and (c)) shows the loss of  orientational order and positional order respectively. 
  \begin{figure*}[!hbtp] 
   \includegraphics[width=4.4cm,height=4.5cm,clip]{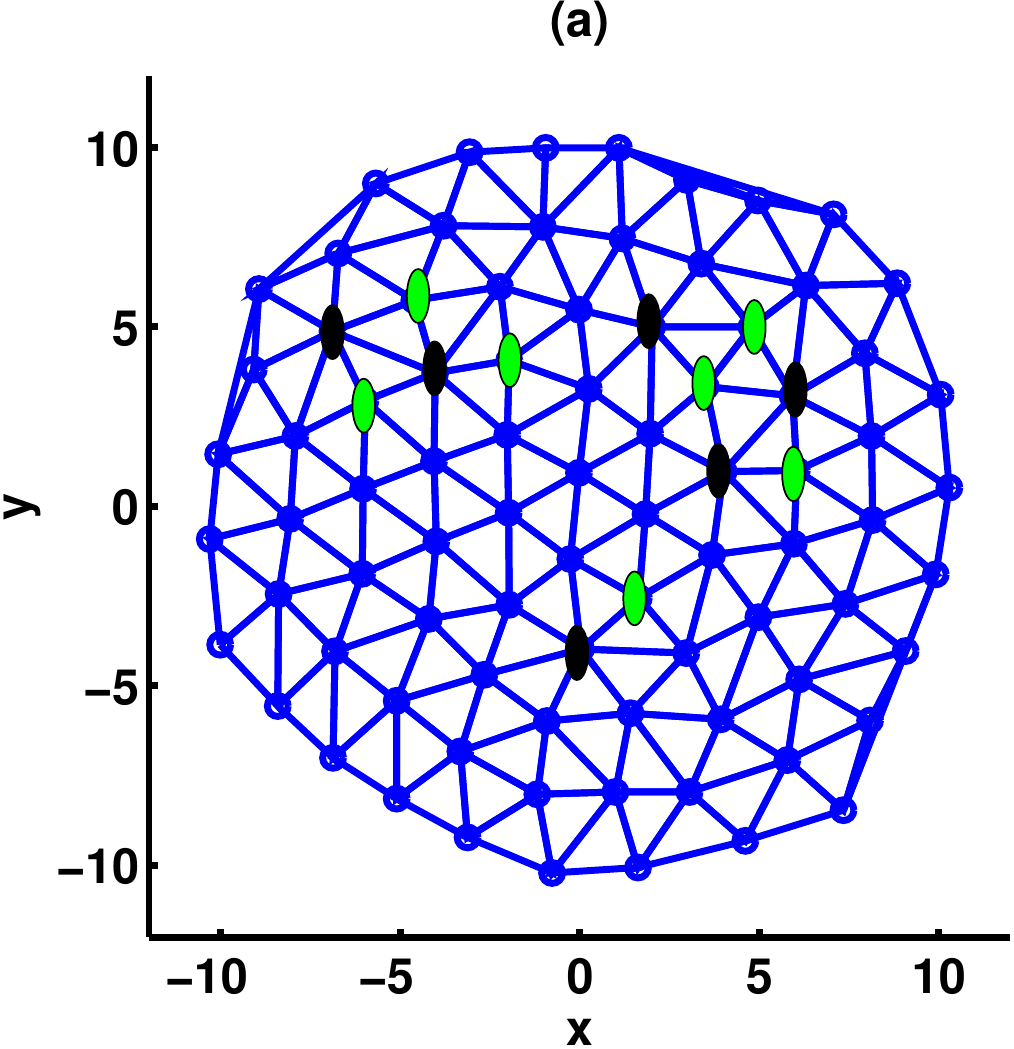}
   \includegraphics[width=4.4cm,height=4.5cm]{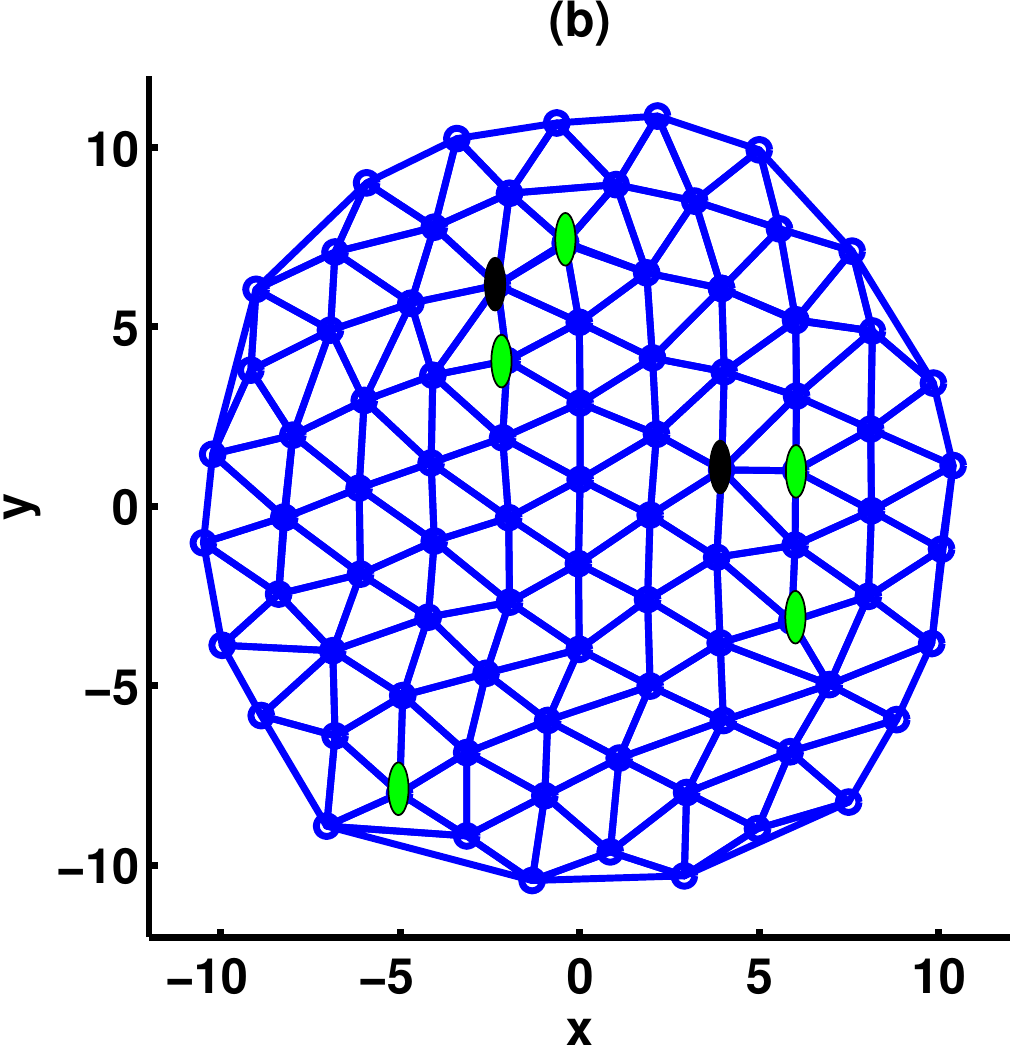}
   \includegraphics[width=4.4cm,height=4.5cm]{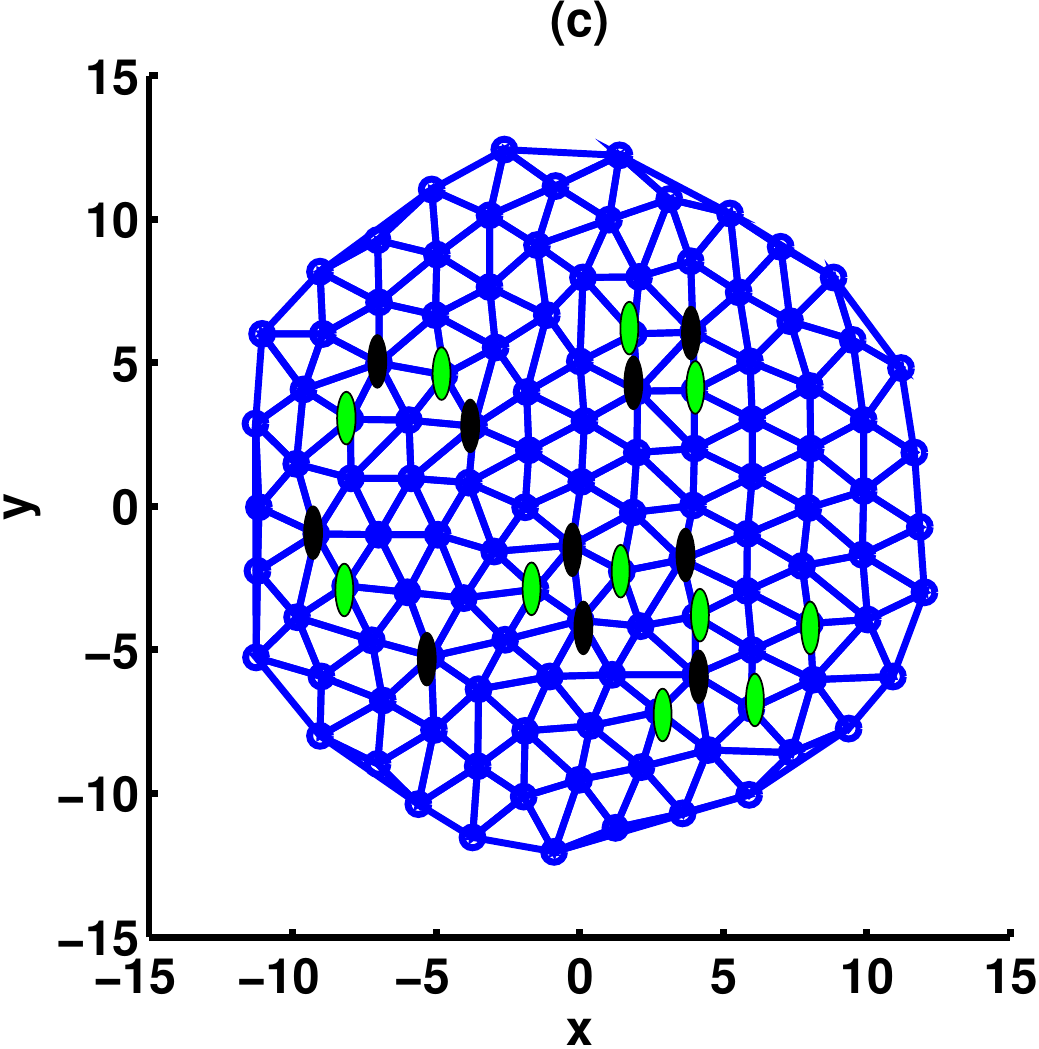}
    \includegraphics[width=4.4cm,height=4.5cm]{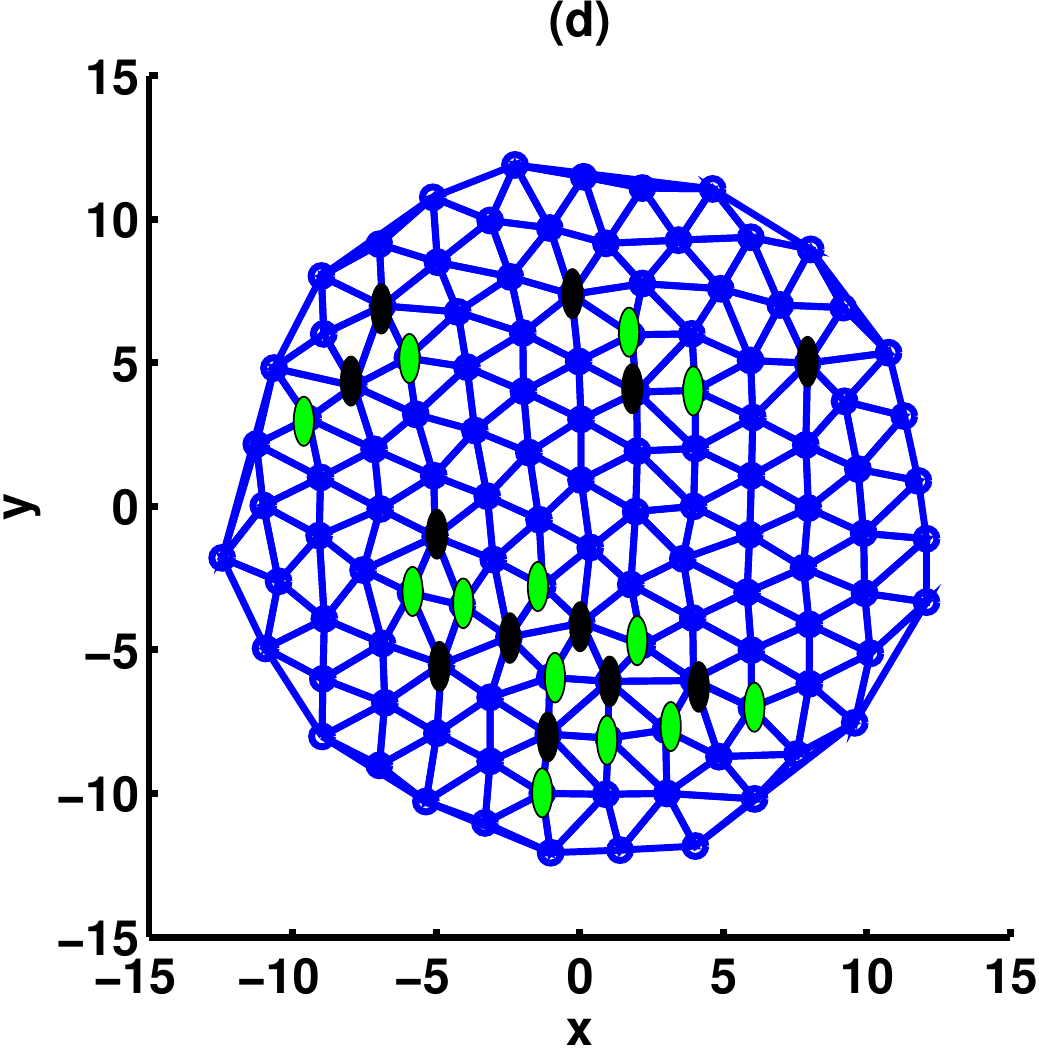}   
        \vspace{-0.8cm}
 \caption{\label{Fig. pulsing_zir}{\footnotesize (Color online) (a) and (c) show Delaunay triangulated disordered vortex lattice at $\Omega=0.7$ and $\Omega=0.8$, respectively  for ZRI case; while, (b) (for \ref{Fig. pulsing_zir}(a)) and (d) (for \ref{Fig. pulsing_zir}(c)) show much ordered lattice after perturbing it with higher rotational frequency. In both cases, initial perturbation involve increasing the rotation frequency by an additional amount 0.05 and then decreasing to their respective original frequencies. Later, the same process is repeated with an additional frequency 0.07. Figures show that the density of defects has not been changed much when compared to RI.}}
      \vspace{-0.4cm}
  \end{figure*}
  \subsection{RI protocol}
 Fig. \ref{Fig. rs} displays the vortex lattice prepared using RI protocol. In this case for the same rotational frequencies, vortex lattice is much 
 more disordered. For $\Omega =0.6$, Fig. \ref{Fig. rs}(a) shows seven dislocations. When compared with corresponding case in ZRI (Fig. \ref{Fig. zrs}
 (a)), it is clear that the state obtained with RI is more disordered against the impurities.  Fig. \ref{Fig. rs}(b) shows that the number of 
 dislocations has increased as similar in ZRI; while, Fig. \ref{Fig. rs}(c) shows the presence of a disclination in addition to the dislocation. 
 Corresponding ZRI case is free from the disclination. The number of disclinations has been increased at higher rotation, $\Omega = 0.95$,  into 
 two as shown in Fig. \ref{Fig. rs}(d). SF shown  Fig. \ref{Fig. rs_sf}(a) and (b) corresponding to cases $\Omega=0.9$ and $0.95$, respectively show the disappearance of both the orientational and positional orders. 
 \subsection{Meta-stable states}
 Further, in order to establish the state obtained using RI protocol is metastable we apply a small perturbation by increasing $\Omega$ momentarily by a small amount and then bringing down to its original value. We observe that the number of defects significantly reduces after this perturbations, Fig. \ref{Fig. pulsing_ir}, confirming that the state prepared in RI protocol is indeed a metastable state, analogous to the FC vortex lattice state in a Type-II supercoductor \cite{Somesh:2016:a}. In contrast, perturbation of similar magnitude does not much alter the states in ZRI as shown Fig. \ref{Fig. pulsing_zir}. Since the states obtained in ZRI protocol are stable against perturbations, these states represent the ground states of the rotating system in presence of random impurities.
 
\section{Conclusion}
 We have shown that the disorder induced vortex lattice melting in BEC follows
a two-step process. This vortex lattice melting process in BEC closely mimics the recently observed disorder-induced two-step melting of vortex matter in weakly pinned type-II superconductor Co-intercalated NbS$e_2$ \cite{Somesh:2016:a}. Using numerical perturbation analysis we have further shown that the vortex lattice states prepared in RI protocol are metastable, analogous to the FC vortex states in type-II superconductors, while the ZRI states are more stable \cite{Somesh:2016:a}. Since the optical lattice potential can easily be modified in a cold atom experiment, our study suggest that a rotating BEC can be powerful tool to investigate variety of vortex lattice states that could emerge in a conventional type-II superconductor under different kind of pinning. This is another analogy, where complex process in type-II superconductors can be studied in the much more accessible condensates. By changing the relative rotation frequency between the condensate and the random optical lattice it should be possible to study about the dynamical phase of vortices.  
\acknowledgments
Authors thank R. Sensarma for useful discussions and V. Tripathi for the helps provided for simulations. T. M., P. R and B. D thank the Science and Engineering Research Board, Government of India, for funding under the schemes National Post-Doctoral Fellowship (Ref. No. PDF/2016/000364), Grant No. EMR/2015/000083 and Grant No. EMR/2016/002627, respectively.

\end{document}